\documentstyle[11pt,IAU207_pasp,twoside,psfig]{article}
\def\edcomment#1{\iffalse\marginpar{\raggedright\sl#1\/}\else\relax\fi}
\reversemarginpar

\begin{document}
\title{Washington Photometry of New Globular Clusters in M31}
\author{Sang Chul Kim, Myung Gyoon Lee}
\affil{Astronomy Program, SEES, Seoul National University\\
   Seoul 151-742, Korea}
\author{Doug Geisler, Juan Seguel}
\affil{Departamento de F{\'\i}sica, Grupo de Astronom{\'\i}a, Universidad
   de Concepci\'on, Casilla 160-C, Concepci\'on, Chile}
\author{Ata Sarajedini}
\affil{Astronomy Department, University of Florida, Gainesville, FL 32611-2055, USA}
\author{William E. Harris}
\affil{McMaster University, Dept. of Physics and Astronomy,
   Hamilton ON L8S 4M1, Canada}

\begin{abstract}
We present a progress report on Washington photometry of several hundred new 
globular cluster (GC) candidates in M31 
which were recently found from our new CCD survey of GCs.
Washington $CMT_1$ filters we used are very efficient 
to survey extragalactic GCs and 
to estimate the metallicity of GCs.
Preliminary color-magnitude diagrams and color-color diagrams of 
the new GC candidates 
and known GCs in M31 are obtained. 
\end{abstract}

M31 is the nearest giant spiral galaxy that 
contains more GCs than our Galaxy and 
all the other Local Group galaxies combined.
Since there has been no CCD survey of GCs in M31
for the entire field of M31 ($> 3^\circ \times 3^\circ$),
the luminosity function of M31 GCs in the present literature
lacks data at the faint end compared to that of our Galaxy. 
Therefore, we have performed a new wide-field CCD survey of 
GCs in M31 over the past few years (Lee et al. 2001).

\begin{figure}
\centerline{
\psfig{figure=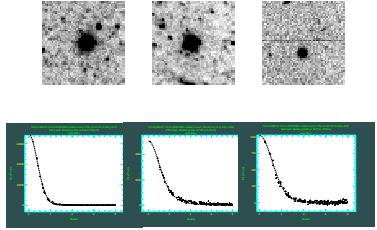,height=6cm} 
}
\vskip -2.4cm
\caption{(Upper panel, from left to right) $T_1$ images 
($34^{\prime\prime} \times 34^{\prime\prime}$) of a star 
($T_1 = 14.3$, $C-T_1=1.2$), 
a known GC (315-038, $V = 16.3$, $B-V=0.4$), and
a new GC candidate ($T_1 = 19.3$, $C-T_1=1.4$).
(Lower panel, from left to right) Radial profiles of 
the star, the known GC, and the new GC candidate.
Note that the FWHM of the star and the typical FWHM of the image is 1.8 px.
The FWHM of the known GC is 2.6 px, and that of the new GC candidate is 2.8 px.}
\end{figure}

Here we present a preliminary Washington photometry 
of new GCs in M31 discovered recently in our new survey.
Our photometry is based on the CCD images obtained using 
the KPNO 0.9m telescope + T2KA $2048 \times 2048$ pixel$^2$ CCD 
(each $23.\arcmin2 \times 23.\arcmin 2$) with Washington $C, M, T_1$ filters.
We have derived the photometry of the objects in the CCD images 
using DAOPHOT II/ALLFRAME.

New GC candidates in the list of objects with photometry 
were selected as follows. 
First, we plotted the color-magnitude and color-color diagrams 
of all the point sources detected in each field, and
took the objects in the regions where GCs are supposed to be
(Harris \& Canterna 1977).
Second, we performed the digital classification using three 
morphological classification parameters (wing moment $r_1$,
central concentration $r_{-2}$, and magnitude difference between PSF fitted magnitude
and aperture magnitude $\Delta$ mag).
Third, we performed visual examination of the images
using the radial/contour profiles and surface plots
of the objects (Fig. 1).
Finally the photometric candidates are to be confirmed spectroscopically.

Figure 2 shows the color-magnitude and color-color diagrams 
of the resulting $\sim 1000$ GC candidates.
From the search processes described above,
we assigned class to each object: 
class 1 for probable GCs, 
class 2 for possible GCs, 
and class 3 for doubtful GCs. 
There are $\sim 100$ class 1 candidates, $\sim 500$ class 2 candidates, 
and $\sim 400$ class 3 candidates in Figure 2.
There are seen more objects in the fainter part ($T_1 > 18$ mag)
than in the brighter part,
probably due to the presence of non-GC objects.
When the objects in this plot are confirmed as genuine GCs
from spectroscopic observations (Lee et al. 2001), 
the fainter part of the luminosity function of M31 GCs
is expected to be filled, 
especially at $18 < V < 20$ mag (i.e. $18.5 < T_1 < 20.5$ mag).

The color-color diagram shows that the new GC candidates (class 1)
are located along the sequence of the known GCs in M31 and our Galaxy,
indicating that most of class 1 objects are probably the genuine GCs in M31.
The large scatter for the M31 GC candidates is expected to be reduced,
when aperture photometry is used instead of the PSF fitting photometry
which was used temporarily in this presentation.
We will finally use aperture photometry of the GCs 
for the analysis of luminosity functions 
and metallicity distribution of M31 GCs.


\acknowledgments
This work was supported in part by the Korea Research Foundation Grant (KRF-2000-DP0450).

\begin{figure}
\centerline{
\psfig{figure=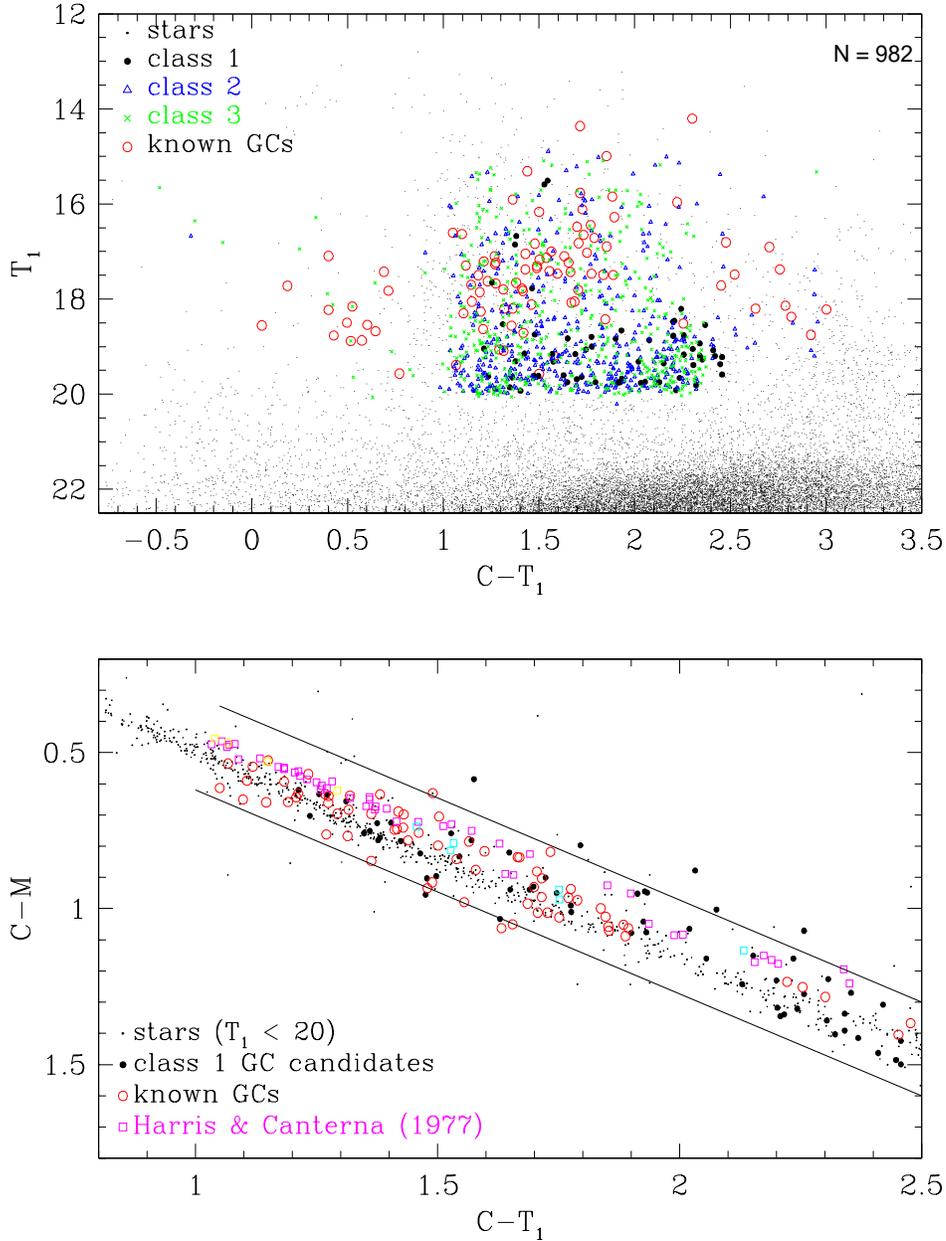,width=13cm,height=18cm}
}
\caption{Color-magnitude diagram (upper panel)
and color-color diagram (lower panel) of new GC candidates.
Stars of one field are plotted as small dots.
Class 1 (probable GCs), class 2 (possible GCs)  and 
class 3 (doubtful GCs) objects are plotted by filled circles,
triangles, and $\times$'s, respectively.
Open circles present $\sim 100$ known GCs in M31.
Open squares in the color-color diagram are GCs in the Galaxy, Fornax dSph,
and M31 taken from Harris \& Canterna (1977).
Objects deviating much from the main locus need further checking.
}
\end{figure}

\end{document}